\documentstyle[12pt]{article}
\def\be{\begin{equation}}
\def\ee{\end{equation}}
\def\ba{\begin{eqnarray}}
\def\ea{\end{eqnarray}}
\def\half{{1\over2}}
\def\ta{\widetilde{a}}

\def\tg{\widetilde{g}}
\def\tp{\widetilde{p}}
\def\tR{\widetilde{R}}
\def\tthreeR{\ ^{(3)}\tR}
\def\trho{\widetilde{\rho}}
\def\tsigma{\widetilde{\sigma}}
\def\tLmatter{\widetilde{\cal L}_{\rm matter}}
\def\tt{\widetilde{t}}
\def\tT{\widetilde{T}}
\def\ttheta{\widetilde{\theta}}
\def\tV{\widetilde{V}}
\def\tnabla{\widetilde{\nabla}}
\newcommand\lsim{\mathrel{\rlap{\lower4pt\hbox{\hskip1pt$\sim$}}
    \raise1pt\hbox{$<$}}}
\newcommand\gsim{\mathrel{\rlap{\lower4pt\hbox{\hskip1pt$\sim$}}
    \raise1pt\hbox{$>$}}}
\begin{document}
\begin{flushright}
SUSSEX-AST-95/1-2\\
gr-qc/9501039\\
\end{flushright}
\bigskip
\begin{center}
\Large
{\bf Anisotropic scalar-tensor cosmologies\\}
\normalsize
\bigskip
{\large Jos\'e P. Mimoso\\}
\bigskip
{\em Dep. F\'isica, F.~C.~L., Ed C1,\\
Campo Grande, 1700 Lisboa, Portugal\\}
\bigskip
{\large David Wands\\}
\bigskip
{\em Astronomy Centre,\\School of Mathematical and Physical Sciences,\\
University of Sussex, Brighton, BN1 9QH, U.K.\\}
\bigskip
{\small PACS nos. 98.80.Hw, 98.80.Cq, 04.50.+h\\}
\end{center}
\bigskip
%\double

\begin{abstract}
%\double
We examine homogeneous but anisotropic cosmologies in scalar-tensor
gravity theories, including Brans-Dicke gravity. We present a method
for deriving solutions for any isotropic perfect fluid with a
barotropic equation of state ($p\propto\rho$) in a spatially flat
(Bianchi type~I) cosmology. These models approach an isotropic,
general relativistic solution as the expansion becomes dominated by
the barotropic fluid.  All models that approach general relativity
isotropize except for the case of a maximally stiff fluid. For stiff
fluid or radiation or in vacuum we are able to give solutions for
arbitrary scalar-tensor theories in a number of anisotropic Bianchi
and Kantowski-Sachs metrics. We show how this approach can also be
used to derive solutions from the low-energy string effective action.
We discuss the nature, and possibly avoidance of, the initial
singularity where both shear and non-Einstein behavior is important.
\end{abstract}

\newpage
\section{Introduction}

Scalar-tensor theories of gravity \cite{Jordan59,Brans+Dicke61,STT}
allow the gravitational coupling to vary, becoming a
dynamical field rather than a fixed constant. This occurs in a range
of fundamental theories that seek to incorporate gravity with the
other interactions. In Kaluza-Klein models this arises from the
variation of the size of the internal dimensions.  In string theories
the dilaton is a scalar field that is necessary for the consistent
description of the motion of a string in a curved spacetime
\cite{effaction}.  In a cosmological context such theories allow one
to seek dynamical answers for questions such as why the Planck mass is
so much larger than other physical scales or why it should appear to
be fixed today.

The general form of the extended gravitational action in scalar-tensor
theories is
\be
S = {1\over16\pi} \int d^4x \sqrt{-g} \left[
		F(\varphi)R - {1\over2}(\nabla\varphi)^2 - U(\varphi)
		+ 16\pi {\cal L}_{\rm matter} \right]
\label{eSvarphi}
\ee
where in general relativity $m_{\rm Pl}^2\equiv F(\varphi)$ remains a
constant.  Brans and Dicke's model of gravity \cite{Brans+Dicke61}
corresponds to the
particular choice of $U=0$ and $F(\varphi)=\varphi^2/8\omega$ where $\omega$
is a constant parameter. The more general Lagrangian can still be
re-written in terms of their Brans-Dicke field $\phi\equiv F(\varphi)$
with the Brans-Dicke parameter $\omega(\phi)=F/[2(dF/d\varphi)^2]$
\be
S = {1\over16\pi} \int d^4x \sqrt{-g} \left[
	\phi R - {\omega(\phi)\over\phi}(\nabla\phi)^2 - U(\phi)
	+ 16\pi {\cal L}_{\rm matter} \right]
\label{eSphi} \ ,
\ee
and it is this form that we shall use here with $U(\phi)$ set to zero
to ensure a strictly Newtonian weak field limit. The post-Newtonian
parameters of general relativity are also recovered in the limit that
$\omega\to\infty$ and $(\phi/\omega^3)(d\omega/d\phi)\to0$ \cite{Will93}.
The case where $U(\phi)$ is non-zero but $\omega=0$ is equivalent to
higher-order gravity theories \cite{Wands94} with Yukawa type
corrections to the Newtonian potential.

Most analytic cosmological solutions until recently were restricted to
the case of Brans-Dicke gravity or a few other specific choices of
$\omega(\phi)$
\cite{Nariai68,GurFinRub73,Nariai72,MatRyaTot73,Bel+Kha73,LP}.
Here we present techniques for deriving anisotropic cosmological
solutions for general scalar-tensor theories where $\omega$ may be an
arbitrary function of $\phi$. This can lead to radically different
evolution of the Brans-Dicke field as well as the Brans-Dicke parameter.
Throughout we draw heavily on results derived in the conformally related
Einstein frame \cite{Dicke62}, introduced in Sect.~(\ref{sectConf}). In
Sect.~(\ref{homog}) we define the quantities we will use to describe the
evolution of homogeneous spacetimes.

We give solutions for barotropic fluids, including dust and a false
vacuum energy density in Sect.~(\ref{sectBI}), extending to Bianchi
type~I models the method used recently by Barrow and Mimoso
\cite{Bar+Mim94} in spatially flat FRW models. This also allows us to
consider the role of general relativity as a cosmological attractor
within a large class of $\omega(\phi)$ theories.

In Sect.~(\ref{sectBx}) we extend the method introduced by Barrow to
derive solutions for Friedmann-Robertson-Walker (FRW) in vacuum or
containing radiation \cite{Barrow93} or stiff matter \cite{Mim+Wan95},
to solve for homogeneous spacetimes including spatial curvature as well
as shear. In particular we give general solutions for arbitrary
$\omega(\phi)$ theories in vacuum, or with radiation or stiff fluid, in
Bianchi types~I and~V, as well as the general vacuum or stiff fluid
solutions in Bianchi types~III and locally rotationally symmetric (LRS)
type~IX and Kantowski-Sachs, by exploiting known solutions in general
relativity.

\section{Conformal frames}
\label{sectConf}

Our field equations, obtained by varying the action in Eq.~(\ref{eSphi}) with
respect to the metric and field $\phi$, are
\ba
\left( R^{ab} - \half g^{ab} R \right) \phi & = &
 8\pi T^{ab}
 + \left( g^{ac}g^{bd}
	- \half g^{ab}g^{cd} \right)
		\frac{\omega\phi_{,c}\phi_{,d}}{\phi}
\nonumber \\
& & \qquad + \left( g^{ac}g^{bd}
			- g^{ab}g^{cd} \right)
		\nabla_{c}\nabla_{d}\phi \label{eEOMg} \\
(3+2\omega) \Box\phi & = & 8\pi T - g^{ab} \omega_{,a}\phi_{,b}
 \label{eEOMphi}
\ea
where the energy-momentum tensor
$T^{ab}=2(\delta{\cal L}_{\rm matter}/\delta g_{ab})$.

In the scalar-tensor gravity theories
the principle of equivalence is guaranteed by requiring that
all matter fields are minimally coupled to the metric $g_{ab}$.
Henceforth we will refer to this as the Jordan metric. Thus
energy-momentum is conserved:
\be
\nabla^{a} T_{ab} = 0 \; .
\ee

However scalar-tensor theories can be
re-written in terms of a theory with a fixed gravitational constant with
respect to the conformally related ``Einstein'' metric:
\be
\tg_{ab} \equiv G\phi g_{ab} \; ,
\ee
where $G$ is in fact an arbitrarily chosen constant which becomes the
gravitational constant in the conformal metric.
Note that if $\phi=$const then the two frames are identical (allowing for
an arbitrary constant rescaling of coordinates) and so the scalar-tensor
results must be the same as in general relativity whenever this occurs.
Notice also that for $\phi<0$ we must pick a negative $G$ to maintain a
positive conformal factor. Henceforth we shall assume $\phi\geq0$.

Instead of appearing in the gravitational Lagrangian, the Brans-Dicke
field now appears as a scalar field interacting with matter.
\be
S = {1\over16\pi} \int d^4x \sqrt{-\tg}
	\left[ {1\over G} \tR - \half {(\tnabla\phi)^2\over\phi^2}
		+ 16\pi \tLmatter \right] \; ,
\ee
where the covariant derivative $\tnabla_a$ is taken with respect to the
Einstein metric and
\be
\tLmatter \equiv {{\cal L}_{\rm matter}\over(G\phi)^2} \; .
\ee
Thus, although we recover the familiar Einstein field equations,
energy-momentum is no longer conserved independently of the Brans-Dicke
field
\be
\tnabla^a \tT_{ab} = - \half {\phi_{,b}\over\phi} \tT_a^a \; ,
\ee
except when the energy-momentum tensor is traceless --- corresponding to
vacuum or radiation. In general it is the difficulty of including this
interaction between matter and the Brans-Dicke field which limits our
ability to produce analytic solutions.

Of course the overall energy momentum tensor is conserved (as
guaranteed by the Ricci identities) as long as we include the Brans-Dicke
field as a matter field with energy-momentum tensor
\be
\hat{T}^{ab} = \left( \tg^{ac}\tg^{bd}
			- \half\tg^{ab}\tg^{cd} \right)
		\psi_{,c}\psi_{,d} \ ,
\label{edefThat}
\ee
where we define
\be
d\psi\equiv \sqrt{{3+2\omega(\phi)\over16\pi G}} \ {d\phi\over\phi} \; .
\label{edefpsi}
\ee
Here we take $3+2\omega\geq0$. \footnote
{If this is not so then the scalar field
in the Einstein frame has a negative energy density and the Minkowski
vacuum, for instance, will be unstable. This seems to be a strong
physical argument for rejecting such models and henceforth we shall
assume $3+2\omega\geq0$.}

This allows us to deal most easily with other massless fields in the
matter Lagrangian, exploiting known solutions in general relativity to
produce scalar-tensor counterparts \cite{Mim+Wan95}. Short-wavelength
modes of a field, $\varphi\propto\exp(ik_ax^a)$ where $k_ak^a=0$, act
like radiation with a traceless energy-momentum tensor. Long-wavelength
modes of a massless scalar field, $\varphi(t)$, act like a stiff fluid
with density equal to pressure equal to $\dot{\varphi}^2/2$
\cite{Tab+Tau73}.  Although this minimally coupled scalar-field in the
Jordan frame interacts with the Brans-Dicke field in the Einstein
frame, they combine to give the same dynamical effect as a that of
single stiff fluid.  We will also consider massless fields which occur
in the low energy effective action of string theory.

We will present in the Sect.~(\ref{sectBx}) results for scalar-tensor
gravity in a number of anisotropic cosmologies in vacuum, with a stiff
fluid and, in some cases, radiation. Before that in
Sect.~(\ref{sectBI}) we will give solutions for other barotropic
fluids but restricted to a Bianchi type~I metric.

\section{Homogeneous spacetimes}
\label{homog}

Spatially homogeneous spacetimes admit a group
of isometries acting transitively on their spacelike hypersurfaces
\cite{MCall 79}.
It is possible to write the metrics of these models as
\begin{equation}
{\rm d}s^2 = -dt^2 + \gamma_{\alpha\beta}(t) {W^\alpha}_a {W^\beta}_b
{\rm d}x^a {\rm d}x^b \; ,
\label{espathommetric}
\end{equation}
where the ${W^a}_\alpha$ (dual to ${W^\alpha}_a$) are the invariant
vector fields of the reciprocal group,
$[{W_\alpha}^a,{W_\beta}^a]={C^\delta}_{\alpha\beta}{W_\delta}^a$. The
structure constants of the isometry group ${C^\delta}_{\alpha\beta}$,
satisfy the Jacobi identities
$C^\alpha_{[\beta\delta}C^\lambda_{\gamma]\alpha}=0$.

These models fall into two classes. Those of Bianchi type and
Kantowski-Sachs models. The former have a three dimensional group
acting in a simple-transitive way on the spatial hypersurfaces. They
separate into equivalence classes according to Bianchi's classification
of the three dimensional Lie groups~\cite{Ellis+MCall 69,MCall 79}.
The latter has a four dimensional isometry group, but no three
simple-transitive subgroup acting on the three space. Instead it acts
on a two-dimensional subspace. This subspace has constant (positive)
curvature.

Although in all spatially homogeneous models the unit normal to
the spatial hypersurfaces is a geodesic vector field $t^a$
(with $t_at^a=-1$) invariant under
the group~\cite{MCall 79}, the matter flow might be tilted relative
to this direction. For the sake of simplicity, in what follows we shall
only be concerned  with those models in which the velocity of matter
is parallel to the unit normal. It follows that
$T^{ab}$ has a timelike eigenvector and, thus, both
rotation and acceleration are vanishing.

The homogeneous three dimensional spatial hypersurfaces thus have a metric
$h_{ab}$, orthogonal to the unit vector field, such that the full
four-dimensional metric in the Jordan frame
\be
g_{ab} = h_{ab} - t_a t_b \ .
\ee
A perfect co-moving fluid, with density $\rho$, isotropic
pressure $p$, and (possibly) anisotropic stress $\pi_a^b$ then has an
energy-momentum tensor
\be
T_a^b \equiv \rho t_at^b + p h_a^b + \pi_a^b\ .
\ee
We will only consider isotropic matter in what follows with
$\pi_a^b=0$ (although one should bear in mind that an anisotropic
expansion can induce anisotropic pressures in some fluids
\cite{Mim+Cra93}). The extrinsic
curvature of the hypersurfaces can be defined as
\be
t_{a;b} \equiv {1\over3}\theta h_{ab} + \sigma_{ab} \ ,
\ee
where the expansion $\theta\equiv t^a_{;a}$ and the shear
$\sigma^2\equiv\sigma_a^b\sigma_b^a/2$. We will also find it
convenient to define a volume scale factor $V$ such that
\be
\theta \equiv {1\over V} V_{,a}t^a \equiv {1\over V} {dV\over dt} \ .
\ee

Note that if we redefine the contribution of the derivatives of the
Brans-Dicke field to the right-hand side of the scalar-tensor equations
of motion in Eq.~(\ref{eEOMg}) to be an effective energy-momentum tensor
$\bar{T}_{ab}$, so that
\begin{equation}
\left( R_{ab} - \half g_{ab}R \right) \phi
 = 8\pi \left( T_{ab} + \bar{T}_{ab} \right) \ ,
\end{equation}
then we find that the non-minimal coupling of the scalar-field to the
spacetime curvature creates an anisotropic stress which is proportional
to the shear:
\begin{equation}
\bar\pi_{ab}^{(\phi)}
 = \frac{1}{8\pi} \frac{\dot{\phi}}{\phi} \sigma_{ab} \ .
\label{edefpiab}
\end{equation}
Thus though the Brans-Dicke field will not produce shear where none exists,
it does exert an anisotropic pressure in the presence of shear even
when the other matter is isotropic. On the other hand in the Einstein
frame the definition in Eq.~(\ref{edefThat}) of the energy-momentum
tensor $\hat{T}_{ab}$, associated with the Brans-Dicke field can exert
no anisotropic stress for a homogeneous field.

It is straightforward to calculate conformally transformed
quantities in the Einstein metric where we find
\ba
\tV    =(G\phi)^{3/2} V\ ,\qquad
& d\tt^2   =(G\phi) dt^2\ ,& \qquad
\tsigma^2={\sigma^2\over (G\phi)} \ , \nonumber \\
\trho    ={\rho\over (G\phi)^2} \ ,& \qquad &
\tp      ={p\over (G\phi)^2} \ .
\ea
The metric field
equations~(\ref{eEOMg}) then yield the familiar constraint equation
\be
\ttheta^2 = 24\pi G \left( \trho + \hat{\rho} \right)
 + 3\tsigma^2 - {3\over2} \tthreeR \; ,
\label{eCON}
\ee
where $\tthreeR$ is the curvature scalar of the hypersurfaces of
homogeneity, and the Raychaudhuri equation
\be
{d\ttheta \over d\tt} + {1\over3}\ttheta^2
 = - 4\pi G \left( 3\tp+3\hat{p}+\trho+\hat{\rho}\right)
   - 2\tsigma^2 \; .
\label{eRAY}
\ee
The total energy-momentum in this frame is conserved:
\be
{d\over d\tt}\left(\trho+\hat{\rho}\right)
	+ \ttheta \left(\trho+\hat{\rho}+\tp+\hat{p}\right) = 0 \; ,
\ee
where we include the density $\hat\rho$ and pressure $\hat{p}$ of
the Brans-Dicke field.

But we are also interested in how the Brans-Dicke field and other
matter evolve separately. The continuity equation for isotropic matter
minimally coupled in the Jordan frame remains simplest in that frame
\be
{d\rho\over dt} + \theta \left( \rho + p \right) =0 \; ,
\ee
which can be integrated for a barotropic fluid with $p=(\gamma-1)\rho$ to give
$\rho\propto V^{-\gamma}$ or
$\trho\propto(G\phi)^{(3\gamma-4)/2}\tV^{-\gamma}$ in the Einstein frame.

The equation of motion for the Brans-Dicke field is then
\be
{d^2\psi\over d\tt^2} + \ttheta {d\psi\over d\tt}
 = -\half \sqrt{16\pi G\over 3+2\omega} \left(3\tp-\trho\right)\; .
\label{psieom}
\ee
This is the one equation of motion in the Einstein frame that still
includes an explicit dependence on the form of $\omega(\phi)$.
In vacuum or with radiation the right-hand side is zero and so the
evolution of both $\psi$ and the volume scale factor $\tV$ in the
Einstein frame {\em is independent of the choice of function
$\omega(\phi)$.} We will show in Sect.~(\ref{sectBx}) that these solutions
for $\tV(\tt)$ also include all the scalar-tensor cosmologies
containing stiff matter.
Analytic solutions for other barotropic fluids but
restricted to spatially flat FRW models have recently been given by
Barrow and Mimoso \cite{Bar+Mim94}. We will now extend this method to the
case of Bianchi type I metrics.

%%%%%%%%%%%%%%%%%%%%%%%%%%%%%%%%%%%%%%%%%%%%%%%%
\section{Barotropic matter in Bianchi type~I}
\label{sectBI}
%%%%%%%%%%%%%%%%%%%%%%%%%%%%%%%%%%%%%%%%%%%%%%%%

We will first consider the case of barotropic fluids (where
$p=(\gamma-1)\rho$), such as dust ($\gamma=1$) or false vacuum
($\gamma=0$), but excluding for the time being the exceptional cases of
stiff fluid ($\gamma=2$) or radiation ($\gamma=4/3$) as well as vacuum.
We will also restrict our analysis to the simplest case of a Bianchi
type~I metric, where the structure constants $C^\alpha_{\beta\gamma}$
all vanish and the spatial curvature is zero. The metric can then be
written as
\be
ds^2 = -dt^2 + a_1^2(t) dx^2 + a_2^2(t) dy^2 + a_3^2(t) dz^2 \; .
\label{eBI}
\ee
which includes the spatially flat FRW metric when $a_1=a_2=a_3$.
The expansion is given by
\be
\theta \equiv {3\over a}{da\over dt} \; ,
\ee
where the volume scale factor
\be
V \equiv a^3 \equiv a_1 a_2 a_3 \; .
\ee

The $\psi$-field equation of motion,
Eq.~(\ref{psieom}), is driven by the barotropic fluid density in the
Einstein frame
\be
\trho = {3\over 8\pi G} {M(G\phi)^{(3\gamma-4)/2} \over \ta^{3\gamma}}\; ,
\ee
where $M=8\pi G\rho a^{3\gamma}/3$ is a constant and $\ta^3\equiv\tV$ is
the volume scale factor in the Einstein frame. This same energy density
drives the evolution of the three scale factors whose individual
expansion rates $\ttheta_i\equiv\dot{\ta}_i/\ta_i$, obey the field
equations
\be
\dot{\ttheta}_i + \ttheta \ttheta_i
 = {3(2-\gamma) \over 2} {M(G\phi)^{(3\gamma-4)/2} \over\ta^{3\gamma}}\; .
\label{thetaieom}
\ee

However the difference between the expansion rates in any two
directions is not driven by the isotropic fluid. The shear is
\be
\sigma^2 \equiv {1\over3} \left( \theta_1^2 + \theta_2^2 + \theta_3^2
 - \theta_1\theta_2 - \theta_2\theta_3 - \theta_3\theta_1 \right) \; ,
\ee
so in the conformally transformed Einstein frame it evolves
like the energy density of a minimally coupled stiff fluid
\be
{d\tsigma^2\over d\tt} + 2 \ttheta \tsigma^2 = 0 \; .
\ee
Thus it behaves exactly like a stiff fluid with density
\be
{\tsigma^2 \over 8\pi G} = {3\Sigma^2 \over 32\pi G\ta^6} \ ,
\ee
where $\Sigma^2$ is a constant, evolving in a flat FRW metric with scale
factor $\ta$.

Note that this general relativistic result is not in general
true in scalar-tensor gravity. Just as a minimally coupled scalar
field in the Jordan frame interacts with the field $\psi$ in the
Einstein frame, so shear, which evolves freely in the Einstein frame,
is coupled to the Brans-Dicke field back in the Jordan frame, and we have
\be
\sigma^2 = \frac{3\Sigma^2}{4(G\phi)^2a^6} \; .
\ee
This is a result of the effective anisotropic pressure
(in the presence of shear) induced by the Brans-Dicke field in the Jordan
frame, but absent in the Einstein frame.

We will use the approach developed by Barrow and Mimoso \cite{Bar+Mim94} for
the scalar-tensor solutions in spatially flat FRW metric, itself an
extension of the method used by Gurevich, Finkelstein and Ruban
\cite{GurFinRub73} for Brans-Dicke gravity.
We introduce the time coordinate $\xi$ defined by
\be
d\tt \equiv
\ta^{3(\gamma-1)} (G\phi)^{(4-3\gamma)/2} \sqrt{\frac{3+2\omega}{3}}
 \; d\xi \ ,
\label{edeftimevar}
\ee
and the variables\footnote
{Note that these variables correspond to those defined by Barrow and
Mimoso in terms of Jordan frame variables where $z\equiv
\sqrt{(3+2\omega)/3}[(x/3)+(y/2)]$ with $G=1$ in that paper\cite{Bar+Mim94}.}
\ba
y &=& \sqrt{\frac{16\pi G}{3}}\
 {\ta}^3 \frac{d\psi}{d\tt} \label{eyEF} \ , \\
z_i &=& \ta^3 \ttheta_i \label{exEF} \ , \\
z &=& {1\over3} \sum_i z_i \ .
\ea

The equations of motion, Eqs.~(\ref{psieom}) and~(\ref{thetaieom}),
reduce to
\ba
y' &=& M\;(4-3\gamma) \ ,
   \label{edy} \\
z_i' &=&
 \sqrt{3+2\omega\over3} {3(2-\gamma)M \over 2}
   \label{edzi} \ ,
\ea
plus the constraint Eq.~(\ref{eCON}), which becomes
\be
4 z^2 = y^2 + \Sigma^2 + 4M (G\phi)^{(3\gamma-4)/2} \ \ta^{3(2-\gamma)} \ .
\label{eFriedm}
\ee
Each term on the right-hand side is non-negative and so we can describe the
expansion in the Einstein frame, $z^2$, at any time as being dominated
by either the term arising from the Brans-Dicke field energy density
$y^2$, the shear $\Sigma^2$, or the matter energy density
$4M(G\phi)^{(3\gamma-4)/2}\ \ta^{3(2-\gamma)}$.

These equations of motion can be integrated to give
\ba
y &=& (4-3\gamma) M (\xi-\xi_*) \ , \label{yint} \\
z_i &=& {3(2-\gamma)M \over 2}
 \left[ \int_{\xi_0}^\xi \sqrt{3+2\omega(\bar\xi)\over3} d\bar{\xi}
	  + \sigma_i \right]
 \ ,
\ea
where $\xi_0$, $\xi_*$ and $\sigma_i$ are integration constants.
Henceforth we will set the constant $\xi_*$ to zero without loss of
generality. This merely amounts to a translation of the time origin.
We have chosen $\xi_0$ so that
\be
z = {3(2-\gamma)M \over 2}
  \int_{\xi_0}^\xi \sqrt{3+2\omega(\bar\xi)\over3} d\bar{\xi}
 \ .
\ee
The integration constants $\sigma_i$ thus characterize the initial
shear and obey
\be
\sum_i \sigma_i =0 \ , \qquad
\sum_i \sigma_i^2 = 2\Sigma^2 \ .
\ee

We can use these results to re-write the equation of motion for the
Brans-Dicke field, Eq.~(\ref{psieom}), solely in terms of $\phi$ and
our time coordinate $\xi$:
\begin{equation}
\left(\frac{\phi'}{\phi}\right)'+\left[ - \frac{4-3\gamma}{2}+
 \frac{(z^2)'}{(4-3\gamma)\xi} \right]\;
\left(\frac{\phi'}{\phi}\right)^2 =
\frac{1}{\xi}\, \left(\frac{\phi'}{\phi}\right)\quad ,
\label{ephi''}
\end{equation}
The solutions to the Bernoulli equation, Eq.~(\ref{ephi''}), can be
cast into the particularly simple form
\begin{equation}
\ln{\left(\frac{\phi}{\phi_0}\right)} = (4-3\gamma)
 \int_{\xi_0}^\xi \frac{\bar{\xi}}{g(\bar{\xi})} d\bar{\xi} \; ,
\label{elnphi}
\end{equation}
where $\phi_0$ is an integration constant, by absorbing
$z(\xi)$ into another function,
\begin{equation}
g(\xi) \equiv {z^2(\xi)\over M^2}
 - \left(\frac{4-3\gamma}{2}\, \xi\right)^2 - g_0
 \ ,
\label{edefg}
\end{equation}
with $g_0$ another integration constant.

Comparing the expression for $\phi'/\phi$ obtained from
Eq.~(\ref{elnphi}) with that from Eq.~(\ref{yint}) shows that
\begin{equation}
g = {(G\phi)^{(3\gamma-4)/2} \ \ta^{3(2-\gamma)} \over M}
 = {(G\phi) a^{3(2-\gamma)} \over M} \ .
\label{eV}
\end{equation}
In other words, $4M^2g(\xi)$ is just the energy density term on the
right-hand side of the constraint Eq.~(\ref{eFriedm}), and the
definition of $g$ in Eq.~(\ref{edefg}) is precisely this constraint
equation, where
\be
g_0 \equiv {\Sigma^2 \over 4M^2} \ .
\ee
Thus $g(\xi)$ must be non-negative.

Note that the behavior of the individual scale factors in the string
frame, $a_i(\xi)$, is given by
\begin{equation}
a_i = \left(\frac{Mg}{G\phi}\right)^{\frac{1}{3(2-\gamma)}}\,
\exp{\left[ - \frac{\sigma_i}{2M}
\int\,\frac{\sqrt{3+2\omega(\bar\xi)}}{g(\bar\xi)}\, d\bar\xi\right]}
\; , \label{ea_i}
\end{equation}
which gives an analytic expression dependent on
our ability to perform the integration in the exponential. Note that we
always have the shear $\sigma^2
\propto (G\phi)^{-2}a^{-6}$, where the average scale factor obeys
\be
a^{3(2-\gamma)} = \left( {Mg \over G\phi} \right)
\label{edefa} \ .
\ee
The shear term, the constant $g_0$, in the constraint
Eq.~(\ref{eFriedm}), or equivalently Eq.~(\ref{edefg}), can only remain
important if both $g$ and $y^2$ remain bounded; the latter case implying
that $\xi^2$ remains bounded. Whenever $\xi\to\infty$ the model
isotropizes.

Note that singularities in the Jordan frame, by which we mean here
points at which the volume scale factor, $a^3$, vanishes, can occur
only if $g\to0$ or $\phi\to\infty$. In fact the latter case can be
shown also to require that $g/z^2\to0$ and thus the dynamical role
played by matter (excluding stiff fluid with $\gamma=2$) at the
singularity is negligible in anisotropic models. Thus we reserve a
discussion of the nature of the singularity until Sect.~(\ref{ssectBs})
where we discuss vacuum and stiff fluid models.

The behavior of the coupling $\omega(\phi)$
which defines the scalar-tensor theory is given in terms of $z(\xi)$ by
\begin{equation}
3+2\omega\left[\phi(\xi)\right] =
 \frac{4}{3(2-\gamma)^2} \; (z')^2 \; .
\label{edefomega}
\end{equation}
The $\omega(\phi)$ dependence is only obtained {\em after} we have
solved for the evolution of $\phi$ and $\omega$ as functions of $\xi$,
if we can invert Eq.~(\ref{elnphi}) to find $\xi(\phi)$.  In practice
a theory is chosen by specifying $g(\xi)$ as a generating function
from which $\phi(\xi)$ follows by Eq.~(\ref{elnphi}), and
$\omega(\xi)$ from Eqs.~(\ref{edefg}) and~(\ref{edefomega}).  We have
$a(\xi)$ from Eq.~(\ref{edefa}), as well as $a_i(\xi)$ from
Eqs.~(\ref{ea_i}), and we can relate our time coordinate $\xi$ to the
proper time in the Jordan frame, $t(\xi)$, from
Eq.~(\ref{edeftimevar}).

\subsection{Brans-Dicke gravity}

Brans-Dicke theory is recovered when $\omega=\omega_0$ is a constant
and we see from Eq.~(\ref{edefomega}) that this implies that
\be
z^2(\xi) = {9(2-\gamma)^2 M^2 \over 4}
	  {3+2\omega_0\over3} (\xi-\xi_0)^2
 \ .
\ee
In the isotropic case where the shear $\Sigma^2=0$,
setting $\xi_0=0$ corresponds to what is often referred to
as the solutions being matter (rather than $\phi$) dominated at early
times. In fact because both $g$ and $y^2$ then evolve as $\xi^2$, the
relative terms on the right-hand side of the constraint
Eq.~(\ref{eFriedm}) are strictly proportional and so this could more
accurately be described as a scaling solution. Clearly all Bianchi
type~I solutions approach this behavior, with $a\propto\xi^\mu$ and
$\phi\propto\xi^\nu$, where
\ba
\mu & = & \frac{2(3+2\omega)(2-\gamma)-2(4-3\gamma)}
			{3(3+2\omega)(2-\gamma)^2-(4-3\gamma)^2}
\ , \\
\nu & = & \frac{4(4-3\gamma)}{3(3+2\omega)(2-\gamma)^2-(4-3\gamma)^2}
\ ,
\ea
as $\xi\to\infty$, giving
Nariai's~\cite{Nariai68} isotropic power-law solutions, $a\propto t^m$
and $\phi\propto t^n$, where
\ba
m & = & \frac{2(3+2\omega)(2-\gamma)-2(4-3\gamma)}
		{3(3+2\omega)\gamma(2-\gamma)-(4-3\gamma)(3\gamma-2)}
\ , \\
n & = & \frac{4(4-3\gamma)}
		{3(3+2\omega)\gamma(2-\gamma)-(4-3\gamma)(3\gamma-2)}
\ ,
\ea
upon integration of the time transformation, Eq.~(\ref{edeftimevar}),
irrespective of the initial shear.

However, the $\xi\to\infty$ limit is never reached (for positive
$\phi$) for $\omega<2(\gamma-5/3)/(2-\gamma)^2$ as in this case
$g\to0$ at finite $\xi$ (and would become negative as $\xi\to\infty$).
Nonetheless this universe does correspond to an infinite proper
lifetime in the Jordan frame, but with a late time expansion driven by
both shear and the Brans-Dicke field density, i.e.,
{\em the model does not isotropize}.

\subsection{Approach to general relativity}

Among the wider class of scalar-tensor gravity theories Brans-Dicke
behavior looks atypical. It only occurs at late times where
$g(\xi)\propto\xi^2$ in the limit $\xi\to\infty$.  Otherwise the
generalized Friedmann constraint equation becomes scalar field
dominated if $g(\xi)<\alpha\xi^2$ for any constant $\alpha$ as
$\xi\to\infty$, or matter dominated if $g(\xi)>\alpha\xi^2$ as
$\xi\to\infty$. In this latter case we see from Eq.~(\ref{edefomega})
that the Brans-Dicke parameter $\omega$ must diverge, and [from
Eq.~(\ref{elnphi})] the Brans-Dicke field $\phi$ tends to a constant
value, in other words {\em we recover the general relativistic
behavior at late times}.

Recently Damour and co-authors \cite{Dam+Nor93,Dam+Pol94} have argued
that the general relativistic limit acts as a cosmological attractor
within the parameter space of more general scalar-tensor gravity
theories. This occurs when the Brans-Dicke parameter diverges. In
the notation favored by Damour and co-authors this corresponds to the
Brans-Dicke field $\phi=F(\varphi)$ as defined in Eq.~(\ref{eSvarphi})
having a local maximum with respect the field $\varphi$ (with
$\phi\neq0$). To see how this emerges in our notation, and in the Bianchi
type~I cosmologies, we will consider the limit
\be
F(\varphi) = \phi_0 - \half k\varphi^2 + {\rm O}(\varphi^3) \; .
\ee
This simply corresponds to a pole in the function $\omega(\phi)$
\be
2\omega(\phi) + 3 =
 {1\over 2k} \ \left( {\phi_0 \over \phi_0-\phi} \right)
 \ \left[ 1 + {\rm O}\left(\sqrt{\phi_0-\phi\over\phi_0}\right) \right]
\; .
\ee

This type of behavior occurs as $\xi\to\infty$ and we consider a
generating function
\be
g(\xi) = g_n \xi^{2n} + {\rm O}(\xi^{2n-1}) \; ,
\ee
with $n>1$. Equation~(\ref{elnphi}) then gives the evolution of the
Brans-Dicke field as
\be
\phi = \phi_0 \left[ 1 - {4-3\gamma \over 2(n-1)g_n\xi^{2(n-1)}}
 + {\rm O}(\xi^{-2n+1}) \right] \; ,
\ee
which, via Eq.~(\ref{edefomega}), gives the above $\omega(\phi)$
behavior for
\be
k = {3(2-\gamma)^2 \over 4(4-3\gamma)} {n-1 \over n^2} \; .
\ee
Note that there is actually a upper bound on
$k\leq3(2-\gamma)^2/16(4-3\gamma)$. This corresponds to the
condition that the Brans-Dicke field is over-damped and approaches
$\phi_0$ monotonically. For larger $k$ it will execute damped
oscillations about $\phi_0$ \cite{Dam+Nor93}.

Using Eq.~(\ref{edefa}) to find the limiting behavior of the average scale
factor and using this to determine the relationship between the time
coordinate $\xi$ and the proper time in the Jordan frame,
Eq.~(\ref{edeftimevar}) , we find (for $\gamma\neq0$)
\ba
\xi & \propto &t^{(2-\gamma)/n\gamma} \ ,\\
a & \propto & t^{2/3\gamma} \ ,\\
{\phi_0-\phi \over \phi_0} & \propto & t^{-2(n-1)(2-\gamma)/n\gamma}
 \propto a^{-3(n-1)(2-\gamma)/n} \ ,
\ea
For $\gamma=0$ both $\xi$ and thus $a$ grow exponentially with respect
to the proper time $t$ leading to de Sitter expansion as $\xi\to\infty$.

Note that the shear must vanish relative to the all the other
terms in the constraint Eq.~(\ref{edefg}) as we approach general
relativity. As $\xi\to\infty$ we find $\rho \propto t^{-2}$
but $\sigma^2 \propto t^{-4/\gamma}$.

\section{Massless fields in anisotropic cosmologies}
\label{sectBx}

We will in this section restrict ourselves only to vacuum or to matter
consisting of the short-wavelength modes (radiation, $\gamma=4/3$) and
long-wavelength modes (stiff fluid, $\gamma=2$) of massless fields
\cite{Mim+Wan95}.  Here we can integrate the equations of motion
in the Einstein frame without specifying $\omega(\phi)$.

For radiation the fluid is conformally
invariant and in this case, as we have seen, a perfect fluid,
i.e.~non-interacting, in the Jordan frame remains a perfect fluid in
the Einstein frame. On the other hand a non-interacting stiff fluid in
the Jordan frame does not remain a perfect fluid in the Einstein
frame, but we can still deal with the dynamics in this case as the
homogeneous Brans-Dicke field also becomes a stiff fluid, $\psi$, and
although there is an interaction between the two components their
combined dynamical effect is the same as that of a single perfect
stiff fluid.  Thus if the stiff fluid in the Jordan frame is a
homogeneous minimally coupled scalar field $\varphi$, we can define
the composite scalar, $\chi$, by
\be
d\chi^2 \equiv d\psi^2 + G\phi d\varphi^2 \; ,
\ee
which obeys the equation of motion for a homogeneous minimally coupled
field
\be
{d^2\chi\over d\tt^2} + \ttheta {d\chi\over d\tt}
 = 0 \; .
\ee
The corresponding energy density in the Einstein frame is
\be
\trho_{\chi} = \half\left({d\chi\over d\tt}\right)^2
 = {3\over8\pi G} \ {A^2 \over 4\tV^2} \; ,
 \label{trhochi}
\ee
where $A$ is a constant of integration.
In the absence of a second field $\varphi$ then we simply have
$\chi\equiv\psi$.

Thus with the conformally transformed matter density and pressure, for
radiation and or stiff fluid in the Einstein frame
\ba
{8\pi G\over3} \ \trho & = & {\Gamma\over\tV^{4/3}} +
{MG\phi\over\tV^2}
\label{edeftrho}
 \; , \\
{8\pi G\over3} \ \tp & = & {\Gamma\over3\tV^{4/3}} + {MG\phi\over\tV^2} \; ,
\ea
where $\Gamma$ and $M$ are non-negative constants, we can obtain the
energy density of the Brans-Dicke field in the Einstein frame as
\be
\hat{\rho} = \hat{p} = \half\left({d\psi\over d\tt}\right)^2
 = {3\over8\pi G} \ {A^2-4MG\phi \over 4\tV^2} \; .
\label{edefhatrho}
\ee
Clearly we require $4MG\phi\leq A^2$ for this to correspond to a
non-negative energy density. Thus in the presence of a stiff fluid,
$M\neq0$, this places an upper bound on the value of the Brans-Dicke
field, or equivalently a lower limit on the effective gravitational
coupling constant in the Jordan frame, $G_{\rm eff}\sim\phi^{-1}\geq
(4M/A^2)G$.

The evolution of the volume scale $\tV(\tt)$ can be given if we can solve
the Einstein equations~(\ref{eCON}) and~(\ref{eRAY}) with total energy
density
\be
{8\pi G\over3} \ \left( \hat{\rho}+\trho \right)
 = {A^2\over 4\tV^2} + {\Gamma\over \tV^{4/3}}
\ee
and pressure
\be
{8\pi G\over3} \ \left( \hat{p}+\tp \right)
 = {A^2\over 4\tV^2} + {\Gamma\over 3\tV^{4/3}} \; ,
\ee
as well as the shear and spatial curvature which will  depend on the
Bianchi class. This Einstein frame solution for radiation and a stiff
fluid, in a particular Bianchi or Kantowski-Sachs metric, is
independent of $\omega(\phi)$ and describes the Einstein frame behavior
for any scalar-tensor theory.

It is only at the final stage that we must specify $\omega(\phi)$ in
order to invert
\ba
\sqrt{16\pi G\over3} \chi(\phi)
 & = &
 \pm \int \sqrt{A^2\over A^2-4MG\phi}\ \sqrt{3+2\omega(\phi)\over3}
	\ {d\phi\over\phi} \ , \label{chiintegral} \nonumber \\
 & = & A \int {d\tt \over \tV} \; , \label{echiint}
\ea
and thus obtain the evolution of $\phi(\chi)$, which is both the
Brans-Dicke field and the conformal factor relating the Einstein frame
to the Jordan metric. We see that the effect of a stiff fluid in the
Jordan frame (i.e., $M\neq0$ for a given value of $A^2$) is to alter the
relation between $\chi$ and $\phi$. But this is the role, in vacuum, solely of
$\omega(\phi)$ and so any scalar-tensor theory, defined by
$\omega(\phi)$, plus a stiff fluid in the Jordan frame is equivalent to
the an effective theory with $\omega_{\rm vac}(\phi)$ in vacuum given
by \cite{Mim+Wan95}
\be
3+2\omega_{\rm vac}(\phi) \equiv {A^2 \over A^2-4MG\phi}\
				\left[3+2\omega(\phi)\right] \ .
\ee
However this equivalence is broken by the presence of any matter other
than radiation (i.e., any matter with non-zero trace of the
energy-momentum tensor) which interacts in the Einstein frame with
the field $\psi$, rather than $\chi$.

Even without knowing the specific form of $\omega(\phi)$ we can note a
few general features of $\chi(\phi)$. In particular if $\chi$ diverges
this must correspond to either $\omega\to\infty$ or
$\ln(G\phi)\to\pm\infty$ (with only $\phi\to0$ possible for $M\neq0$).
The former case may occur as $\phi$ approaches a finite, non-zero
constant value in which case the the Jordan frame coincides (up to an
arbitrary constant factor) with the Einstein frame and we must have
general relativistic behavior. The latter case corresponds to singular
behavior of the conformal factor so we may expect radically different
behavior in the Jordan frame from that in the Einstein frame. On the
other hand as $\chi\to$const we must have either $\omega\to-3/2$ or
$\phi\to$constant. Thus for any $\omega>-3/2$ we must also recover
general relativistic behavior in this limit.

\subsection{Scalar-tensor theories}

We will give here only three particular examples for which we can
perform the inversion to find $\phi(\chi)$ analytically in
Eq.~(\ref{chiintegral}), though we
can solve for any $\omega(\phi$) using numerical integration.

\subsubsection{Brans-Dicke gravity}

When $\omega=\omega_0=$constant we find
\ba
\sqrt{16\pi G\over3} \chi(\phi) & = & \pm \sqrt{3+2\omega_0\over3} \ln G\phi
 \qquad {\rm for\ }M=0 \\
 & = & \pm \sqrt{3+2\omega_0\over3} \ln
  \left[ {|A|+\sqrt{A^2-4MG\phi} \over |A|-\sqrt{A^2-4MG\phi}} \right]
 \quad {\rm for\ }M\neq0 \; .
\ea
Notice that while the presence of a stiff fluid leaves the evolution of
the scale factor in the Einstein frame unaltered (i.e.~independent of
the value of $M$ for a given $A^2$) it affects the form of $\chi(\phi)$
and thus the evolution of $\phi$ and the conformal transform back to the
Jordan frame.

Inverting these relations gives
\ba
G\phi(\chi) & = & \exp\left(\beta_0\sqrt{16\pi G\over3}\chi\right)
 \qquad {\rm for\ }M=0 \\
& = & \frac{A^2}
	{4M\cosh^2\left({\beta_0\over2}\sqrt{16\pi G\over3}\chi\right)}
 \qquad {\rm for\ }M\neq0
\ea
as shown in Fig.(1), where
\be
\beta_0= \pm \sqrt{3\over3+2\omega_0} \ . \label{beta0}
\ee
Thus at an initial singularity in the Einstein frame where
$\chi\to\infty$ we must have $\phi\to0$ or $\phi\to\infty$ for $M=0$.
In the presence of a stiff fluid ($M\neq0$) only $\phi\to0$ is
possible.

\subsubsection{$\omega(\phi)$ with Brans-Dicke and G.R. limits}

A alternative choice of $\omega(\phi)$ that displays both a
Brans-Dicke regime and general relativistic behavior is
\be
2\omega(\phi)+3 = (2\omega_0+3) {\phi^2\over(\phi-\phi_0)^2} \ .
\label{defBDGR}
\ee
Clearly for $\phi\geq\phi_0$ we have $\omega\geq\omega_0$ and
$\omega$ approaches this lower limit as $\phi\to\infty$, while as
$\phi\to\phi_0$ we find $\omega\to\infty$.

Equation~(\ref{chiintegral}) then yields
\ba
\sqrt{16\pi G\over3} \chi(\phi)
 & = & \pm \sqrt{2\omega_0+3\over3} \ln \left( {\phi-\phi_0\over\phi_0}
				      \right) \qquad {\rm for} \ M=0 \ ,\\
 & = & \pm \sqrt{2\omega_0+3\over3} \sqrt{A^2\over A^2-4MG\phi} \nonumber
\\
 & & \qquad \  \ln \left( \frac{ \sqrt{A^2-4MG\phi_0}+\sqrt{A^2-4MG\phi} }
			{ \sqrt{A^2-4MG\phi_0}-\sqrt{A^2-4MG\phi} } \right)
 \quad {\rm for} \ M\neq0 \ .
\ea
Note that $\chi\to\infty$ as $\phi\to\phi_0$, and also as
$\phi\to\infty$ when $M=0$. But because $\phi\leq A^2/4MG$ in the
presence of the stiff fluid the latter Brans-Dicke limit cannot be
reached.

This can be inverted to give
\ba
\phi(\chi)
 & = & \phi_0 \left[ 1 + \exp \left( \sqrt{16\pi G\over 3}\beta_0\chi
		      \right) \right]
  \qquad {\rm for} \ M=0\ , \\
 & = & {A^2\over4MG} - \left({A^2-4MG\phi_0\over4MG}\right)
		       \tanh^2 \left(\half B_0 \sqrt{16\pi G\over3}
				 \chi \right)
  \quad {\rm for} \ M\neq0\ ,
\ea
where $\beta_0$ was defined in Eq.~(\ref{beta0}) and
\be
B_0 = \beta_0 \sqrt{A^2-4MG\phi_0\over A^2} \ .
\ee
The function $\phi(\chi)$ is plotted in Fig.(1).

\subsubsection{Axion-dilaton string cosmologies}

While the effective action derived in the low energy limit of string
theory is sometimes referred to as Brans-Dicke gravity with $\omega=-1$,
this is strictly only true when all other matter fields are minimally
coupled to the Jordan, or `string', metric. This is generally not true
in string theory, but we will show that at least for other massless
fields such as the antisymmetric tensor field, $H_{\mu\nu\lambda}$, and
moduli fields, $\beta$, appearing in the low energy effective action,
the techniques developed for scalar-tensor gravity can also be applied to
string theory. (For a more detailed discussion in the case of isotropic
FRW cosmologies see \cite{CopLahWan94}.)

The background field equations of motion reduced to four dimensions can
be derived from the action \cite{effaction}
\be
S = {1\over16\pi G} \int d^4x \sqrt{-g} e^{-\varphi}
 \left[ R + g^{\mu\nu}\varphi_{,\mu}\varphi_{,\nu}
  - ng^{\mu\nu}\beta_{,\mu}\beta_{,\nu}
  - {1\over12}H_{\mu\nu\lambda}H^{\mu\nu\lambda} \right]
\ .
\ee
Conventionally the variable gravitational coupling is represented by the
dimensionless dilaton field $\varphi$, which is simply related to the
Brans-Dicke field $\phi\equiv1/(Ge^\varphi)$.
If the antisymmetric tensor field is a function only of the
4-dimensional spacetime coordinates then it can be described by a single
(pseudo-)scalar ``axion'' field $h$, where
\be
H^{\mu\nu\lambda} = e^{\varphi} \epsilon^{\mu\nu\lambda\kappa}
h_{,\kappa}
\ ,
\ee
and $\epsilon_{\mu\nu\lambda\kappa}$ is the antisymmetric volume form.

Conformally transforming to the Einstein frame the above action then becomes
\be
S = {1\over16\pi G} \int \sqrt{-\tg}
 \left[ \tR - \half \tg^{\mu\nu}\varphi_{,\mu}\varphi_{,\nu}
  - n \tg^{\mu\nu}\beta_{,\mu}\beta_{,\nu}
  - \half \tg^{\mu\nu} e^{2\varphi} h_{,\mu} h_{,\nu} \right]
\ .
\ee
Thus we see that the dilaton is simply related to our field $\psi$,
defined in Eq.~(\ref{edefpsi}), as $d\psi\equiv-d\varphi/\sqrt{16\pi
G}$. For homogeneous fields we can also define the composite scalar
field $\chi(\varphi)$ such that
\be
d\chi^2 \equiv d\psi^2 + \sqrt{n\over 8\pi G} d\beta^2
 + {e^{2\varphi} \over 16\pi G} dh^2 \; ,
\ee
whose total energy momentum tensor is conserved.
Although the axion field in particular is coupled directly to the
dilaton, we can still integrate the equations of motion to deduce the
relative energy densities in the Einstein frame:
\ba
\trho_\chi & = & \half \left( {d\chi\over d\tt} \right)
 = {3\over 8\pi G} {A^2\over 4\tV^2} \ , \\
\trho_\beta & = & {n\over 16\pi G} \left( {d\beta\over d\tt} \right)
 = {3\over 8\pi G} {B^2\over 4\tV^2} \ , \\
\trho_h & = & {1\over 32\pi G} \left( {dh\over d\tt} \right)
 = {3\over 8\pi G} {M e^{-2\varphi} \over \tV^2} \ .
\ea
and thus
\be
\hat\rho = \half \left( {d\psi\over d\tt} \right)
 = {3\over 8\pi G} {A^2-B^2-4Me^{-2\phi} \over \tV^2} \; .
\ee
Once again we see that the presence of other massless fields places an
upper bound on the value of the Brans-Dicke field, or equivalently a
lower bound on the dilaton,
\be
e^{-\varphi} \leq \sqrt{A^2-B^2 \over 4M} \ .
\ee

The relation between $\chi$ and $\varphi$ is then
\ba
\sqrt{16\pi G \over 3} \chi(\varphi)
& = &
 {1\over2\sqrt{3}} \int {A\over\sqrt{A^2-B^2-4Me^{-2\varphi}}}\
  d\varphi \ , \\
& = & {1\over2\sqrt{3}} \sqrt{A^2 \over A^2-B^2} \ \varphi
 \quad {\rm for}\ M=0\ , \\
& = & {1\over2\sqrt{3}} \sqrt{A^2 \over A^2-B^2} \nonumber \\
& & \ \times \ln \left( {\sqrt{A^2-B^2}+\sqrt{A^2-B^2-4Me^{-2\varphi}}
             \over \sqrt{A^2-B^2}-\sqrt{A^2-B^2-4Me^{-2\varphi}}}
     \right)
\quad {\rm for}\ M\neq 0\ .
\ea
Inverting this gives
\ba
e^\varphi(\chi)
& = & \exp
 \left( 2\sqrt{3} \sqrt{A^2-B^2\over A^2} \sqrt{16\pi G\over3} \chi \right)
\quad {\rm for}\ M=0\ , \\
& = & \left[ {2M\over A^2-B^2}
       \cosh \left( \sqrt{A^2-B^2\over A^2} \sqrt{64\pi G} \chi
              \right) \right]^{1/2}
\quad {\rm for}\ M\neq0\ .
\ea
For $M=0$ and $B=0$ this coincides with the Brans-Dicke result with
$\omega=-1$. For $M=0$ but $B\neq0$ the effective Brans-Dicke parameter
lies in the range $-3/2<\omega<-1$. In the presence of the antisymmetric
tensor field ($M\neq0$) the form of $\phi\equiv (Ge^\varphi)^{-1}$
differs from a purely Brans-Dicke result.

Thus we can use the same results as we will use for general scalar-tensor
gravity models with stiff fluid (with or without radiation) in the
Einstein frame to derive the general solutions to the low energy string
effective action including homogeneous antisymmetric tensor and moduli
fields (with or without radiation).

\subsection{Anisotropic models}
\label{ssectBs}

\subsubsection{Bianchi type~I}
\label{ssectBI}

Considering again the spatially flat, anisotropic metric given in
Eq.~(\ref{eBI}),
we will introduce the variable $X\equiv\ta^2\equiv (G\phi) a^2$ and the
conformally invariant time coordinate $d\eta\equiv d\tt/\ta\equiv
dt/a$. Remembering that the shear in the Einstein frame is given by
$\tsigma^2 = 3\Sigma^2/4\ta^6$, and using $\hat{\rho}$ as given in
Eq.(\ref{edefhatrho}), so that
\be
{\phi' \over \phi}
 = \sqrt{3\over 3+2\omega} \ {\sqrt{A^2-4MG\phi}\over X}
\ ,
\label{edefphiprime}
\ee
and $\trho$ given in Eq.(\ref{edeftrho}), we find the
constraint Eq.~(\ref{eCON}), in vacuum or with a stiff fluid in the
Jordan frame (but for $\Gamma=0$), can be simply written as
\be
X'^2 = A^2 + \Sigma^2 \; .
\ee
This is precisely the same constraint equation as solved in the case of
scalar-tensor gravity in flat FRW models in vacuum \cite{Barrow93} or
with stiff fluid \cite{Mim+Wan95} and we have
\be
X = \ta^2 = \sqrt{A^2 + \Sigma^2} \left|\eta-\eta_0\right| \; .
\ee
In the presence also of radiation in the Jordan frame the constraint
Eq.~(\ref{eCON}) becomes
\be
X'^2 = A^2 + \Sigma^2 + 4\Gamma X \; ,
 \label{eXprime}
\ee
and so
\be
X = \ta^2 = \left|\eta-\eta_0\right| \left( \sqrt{A^2+\Sigma^2}
	  + \Gamma\left|\eta-\eta_0\right| \right) \; .
\ee

These are well-known general relativistic results for the evolution of
Bianchi type I models in the presence of radiation and stiff fluid. As one
would expect the stiff fluid and shear dominate the evolution near the
singularity ($\ta\to0$) but the radiation term will dominate as
$\ta\to\infty$.

The behavior of this averaged scale factor, together with the shear
\be
\tsigma^2 = {3\Sigma^2\over
	     4|\eta-\eta_0|^3(\sqrt{A^2+\Sigma^2}+\Gamma|\eta-\eta_0|)^3}
\; ,
\ee
describe the general evolution. However the Bianchi type~I metric has
three degrees of freedom so there is a degeneracy within the evolution
we have described so far depending on how the expansion and shear is
divided between the three scale factors. Solving the equations of motion
for each scale factor in the Einstein frame we find
\be
\ta_i = |\eta-\eta_0|^{3c_i/2}
	\left(\sqrt{A^2+\Sigma^2}+\Gamma|\eta-\eta_0|\right)^{1-(3c_i/2)}
\; , \label{aBI}
\ee
where the definitions of the overall expansion and shear give two
constraints on the three new integration constants:
\be
\sum_i c_i = 1 \; , \qquad \sum_i c_i^2 = 1 - {2A^2\over 3(A^2+\Sigma^2)} \; .
 \label{BIci}
\ee
Thus although there is a unique (isotropic) late-time behavior in the
the presence of radiation ($\Gamma\neq0$), where
$\ta_i\propto|\eta-\eta_0|$, the initial behavior (for
$|\eta-\eta_0|\ll\sqrt{A^2+\Sigma^2}/\Gamma$) is dependent upon
the choice of integration constants, $c_i$. [See Fig.(2)].

This early power-law evolution of the scale-factors includes the pure
general relativistic ``Kasner'' vacuum solution with $A=0$. Although
the averaged scale factor, $\ta$, always vanishes at $\eta_0$, this
singularity can be point-like (all $c_i>0$ for all $i$) or linear (only
one $c_i<0$) unlike the Kasner behavior for which only the linear
singularity is possible (except when $c_j=1$ and $c_{i\neq j}=0$ and
the singularity is planar). Indeed for $A^2>3\Sigma^2$ only the
point-like singularity is possible \cite{Bel+Kha73}.

To recover the full scalar-tensor results in the original Jordan frame we
must calculate $\chi$ as a function of time from Eq.~(\ref{trhochi}),
\ba
\sqrt{16\pi G\over 3} \left( \chi(\eta)-\chi_0 \right)
 & = & {A\over\sqrt{A^2+\Sigma^2}} \ln |\eta-\eta_0|
	\qquad {\rm for\ }\Gamma=0 \\
 & = & {A\over\sqrt{A^2+\Sigma^2}}
     \ln {\Gamma|\eta-\eta_0| \over \sqrt{A^2+\Sigma^2}+\Gamma|\eta-\eta_0|}
	\quad {\rm for\ }\Gamma\neq0
\ea
and then use $\phi(\chi)$, dependent on the form of $\omega(\phi)$, to
specify the evolution of the Brans-Dicke field, and thus the conformal
factor relating these solutions back to the Jordan frame variables.
Note that the field $\chi$ must always diverge as $\eta\to\eta_0$. (In
the absence of radiation it also diverges at $\eta\to\pm\infty$.)

As noted earlier, the definition of $\chi(\phi)$ in Eq.(\ref{echiint})
requires either $\omega\to\infty$ or $\ln(G\phi)\to\pm\infty$. The former
case where $\phi$ tends to a finite, non-zero value of will lead to a
purely general relativistic
result as $X\to0$. In the latter case we can expand about the point
$X\to0$ and we find that in the Jordan frame we have
\ba
\phi \propto |\eta-\eta_0|^n \ , \\
a \propto |\eta-\eta_0|^{(1-n)/2} \ , \\
a_i \propto |\eta-\eta_0|^{(3c_i-n)/2} \ , \\
t \propto |\eta-\eta_0|^{(3-n)/2} \ ,
\ea
if $\omega\to\omega(0)\neq-3/2$ and $n\neq3$, where from
Eq.(\ref{edefphiprime})
\be
n = \pm \sqrt{3\over 3+2\omega} \ \sqrt{A^2\over A^2+\Sigma^2} \ .
\ee

Thus without specifying the full $\omega(\phi)$ we can deduce a number of
features of the possible behavior in the Jordan frame:
\begin{enumerate}
\item In the presence of shear and/or a stiff fluid in the Jordan
frame, a ``bounce'' (where the volume factor $V$ is stationary,
$dV/dt=0$) will occur whenever
\be
{da\over dt} = \half \left( {X'\over X} - {\phi'\over \phi} \right) = 0
 \ ,
\ee
which, from Eqs.~(\ref{edefphiprime}) and~(\ref{eXprime}), requires
\be
\omega = -{3\over 2} + {3\over 2} \left( {A^2-4MG\phi\over A^2+\Sigma^2}
 \right) \ . \label{Cbounce}
\ee
We thus see that a bounce is only possible when
\be
-{3\over2} \leq \omega \leq 0 \ .
\ee
In a flat FRW metric ($\Sigma^2=0$) we require $\omega=0$
\cite{Mim+Wan95}. The presence of shear or a stiff fluid requires a
negative value of $\omega$, producing a negative effective energy
density for the Brans-Dicke field in the Jordan frame, to produce a
bounce. For $\omega(\phi)$ less than that given in Eq.(\ref{Cbounce}),
the averaged scale factor in the Jordan frame actually grows as both $X$
and $\phi$ approach zero.
\item The singularity at $X=0$ is always present in the Einstein
frame, and is bound to be anisotropic for $\Sigma^2\neq0$, with
$\tsigma^2\to\infty$ as $X\to0$. However we
noted in Sect.~(\ref{homog}) that in the Jordan frame the Brans-Dicke
field exerts an
anisotropic pressure proportional to the shear and one might wonder
whether it is possible for this to suppress the shear as one
approaches the singularity. This is indeed possible. The shear in the
Jordan frame in a Bianchi type~I model is given by
\be
\sigma^2 = {3\Sigma^2 G\phi \over 4X^3} \ ,
\ee
so if $\phi$ vanishes faster than $X^3$ the anisotropic initial
singularity in the Einstein frame becomes isotropic in the Jordan
frame. Again from Eqs.~(\ref{echiint}) and~(\ref{eXprime}) we find
that this requires $\phi\to0$ and
\be
\omega < -{3\over2} + {1\over6}\left(A^2\over
A^2+\Sigma^2\right) \ . \label{Cisotropy}
\ee
\item Is it then possible that the singularity as $X\to0$ in the
Einstein frame can be avoided completely in the Jordan frame? First
note that the condition that $da/dt=0$ [Eq.~(\ref{Cbounce})] is incompatible
with the condition that the shear should vanish [Eq.~(\ref{Cisotropy})], or
even remain finite, as $X$ and $\phi$ both approach zero.
However, the expansion $\theta\equiv3(da/dt)/a$ can
remain finite
even though $da/dt$ diverges if the average scale factor $a$ grows
fast enough. Using the above results for a general scalar-tensor theory
in a Bianchi type~I metric with a stiff fluid, we can write the expansion as
\be
\theta^2 = {3(A^2+\Sigma^2)\over4} \; {G\phi\over X^3} \;
 \left( 1 \pm \sqrt{3\over3+2\omega}\sqrt{A^2-4MG\phi\over
A^2+\Sigma^2} \right) \ .
\ee
We see that whenever $\phi\to0$ as $X\to0$ such that the
shear $\sigma^2\to0$, then the expansion will also vanish (for
$\omega\neq-3/2$). This also coincides with the case where
$\eta\to\eta_0$ takes an infinite proper time in the Jordan
frame. Thus although these models reach an anisotropic singularity in a
finite time in the Einstein frame, {\em this corresponds to a non-singular,
shear-free infinite proper lifetime in the Jordan frame}. The condition
for this to occur is simply the condition for the shear to vanish as
$X\to0$ given above, which can occur for
\be
-{3\over2} < \omega < -{4\over3} \ ,
\ee
in a flat FRW metric ($\Sigma^2=0$), or for the more limited range given in
Eq.~(\ref{Cisotropy}) depending on the relative strength of the
anisotropy.\footnote{The requirement that $\omega<-4/3$ to avoid the
singularity
was pointed out by Nariai \cite{Nariai72} in the case of Brans-Dicke
gravity.}
Remember that $\phi=0$ coincides with $\chi(\phi)\to\infty$ for
$\omega\neq-3/2$ and so $\phi\to0$ can be an attractor solution as
$X\to0$ where we have shown $\chi$ must diverge.
\item If $\phi$ grows sufficiently rapidly as we approach $X\to0$
then the dynamical effect of other matter,
neglected here, may no longer remain negligible. The condition for the
density of isotropic matter with barotropic index $\gamma$ to decrease with
respect to the shear or Brans-Dicke field density in the Einstein
frame as $\eta\to\eta_0$ is that
\be
\omega > -{3\over2} + {3\over2}\left({4-3\gamma\over2-\gamma}\right)^2
\left(A^2\over A^2+\Sigma^2\right) \ .
\ee
For smaller values of $\omega$ it may no longer be possible to
neglect the effect of this matter as $\ta\to0$. In particular in the
case of the non-singular solutions given above with vanishing shear and
expansion (in the Jordan frame) as $\eta\to\eta_0$, the relative
density of barotropic matter always grows as we approach $\eta_0$ for
fluids with $\gamma\leq1$. We have also neglected here any possible
anisotropy in the matter content. This, of course, may well play an important
role in anisotropic solutions, but is beyond the scope of this
paper and we leave this for future work.
\end{enumerate}

\subsubsection{Bianchi type~V}

The next simplest case to consider is that of a Bianchi type~V metric
which can be written as
\be
ds^2 = -dt^2 + a^2(t) \left( dx^2
		+ e^{2x}\left[L^2(t) dy^2 + L^{-2}(t) dz^2 \right] \right)
\ee
Just as Bianchi type~I includes the flat FRW model, Bianchi type~V
includes the open FRW model in the isotropic limit, $L=$const.
Here we have already introduced the averaged scale factor $a$, so that the
expansion is simply
\be
\theta = {3\over a} {da\over dt} \ ,
\ee
and the shear is then
\be
\sigma = {1\over L} {dL\over dt} \ .
\ee
Like open FRW models, the homogeneous hypersurfaces have a
negative spatial curvature
\be
^{(3)}R= {3\over a^2} \; ,
\ee
but because this is a function of the averaged scale factor, i.e.~does
not select out a particular direction, it does not drive the shear.

Thus transforming to the Einstein frame where the Brans-Dicke field
decouples from the spacetime curvature, we find that the shear again evolves
as a free field, $\tsigma^2=3\Sigma^2/4\ta^6$,
just as in Bianchi type~I.
Introducing $X\equiv\ta^2\equiv G\phi a^2$ and
$d\eta\equiv d\tt/\ta\equiv dt/a$ as before, the
constraint equation, including stiff fluid and radiation, simply becomes
\be
X'^2 - 4 X^2 = A^2 + \Sigma^2 + 4\Gamma X \; ,
\ee
This is mathematically identical to the constraint equation solved
in the case of scalar-tensor gravity for an open FRW model
\cite{Mim+Wan95}.
We can integrate this to obtain
\be
X = \ta^2 = \frac{\tau \left(\sqrt{A^2+\Sigma^2}+\Gamma\tau\right)}
		 {1-\tau^2} \; ,
\ee
where we have written
\be
\tau \equiv \tanh |\eta-\eta_0| \; .
\ee
This variable, $\tau$, turns out to play much the same role in the presence
of negative spatial curvature as the conformal time, $\eta$, does in the
spatially flat case, as is the case in FRW models \cite{Mim+Wan95}.

At late times ($\eta\to\pm\infty$) $\tau\to1$ and the evolution
becomes curvature dominated, as we would expect in a non-inflationary
universe. Because the curvature does not drive the shear,
\be
\tsigma^2 = {3(1-\tau^2)^3\Sigma^2\over
	     4\tau^3\left(\sqrt{A^2+\Sigma^2} +\Gamma\tau\right)^3} \; ,
\ee
it vanishes as $\tau\to1$.

At early times $\tau\simeq|\eta-\eta_0|\ll 1$ and so the
curvature is irrelevant and we recover
a Bianchi type~I solution. However unlike the general Bianchi type~I,
the metric has only two degrees of freedom
and so its evolution in the Einstein frame is
completely described by the expansion and shear.
Integrating the expression for $\tsigma$ gives
\ba
L^2 & = & L_0^2 \tau^{\sqrt{3}\Sigma/\sqrt{A^2+\Sigma^2}}
  \qquad {\rm for}\ \Gamma=0 \; , \\
 & = & L_0^2 \left({\tau\over\sqrt{A^2+\Sigma^2}+\Gamma\tau}
	     \right)^{\sqrt{3}\Sigma/\sqrt{A^2+\Sigma^2}}
  \qquad {\rm for}\ \Gamma\neq0 \; .
\ea
The behavior near the singularity in Bianchi type~V metrics is thus
only a subset of the Bianchi type~I solutions, given in
Eqs.~(\ref{aBI}) and~(\ref{BIci}), restricted to
$c_2-c_3=\sqrt{3}\Sigma/(A^2+\Sigma^2)$. Thus (for a given $A$ and
$\Gamma$) they are parametrized solely by the choice of $\Sigma$.

Similarly, because the Brans-Dicke field is decoupled from the
spacetime curvature in the Einstein frame, the field
\ba
\sqrt{16\pi G\over3} (\chi-\chi_0) & = & {A\over\sqrt{A^2+\Sigma^2}} \ln \tau
 \qquad {\rm for}\ \Gamma=0 \; ,\\
& = & {A\over\sqrt{A^2+\Sigma^2}}
      \ln {\Gamma\tau\over \sqrt{A^2+\Sigma^2}+\Gamma\tau}
 \qquad {\rm for}\ \Gamma\neq0 \; ,
\ea
and thus $\phi(\chi)$ approaches a fixed value $\phi_\infty$ as
$\tau\to 1$ (unless $\chi(\phi)$ is singular at this point).
Thus the evolution of the metric in the original Jordan frame will
also approach that in the Einstein frame at late times.

\subsubsection{Bianchi type~III and Kantowski-Sachs}

Here we will write the metric in the Jordan frame as
\be
ds^2 = -dt^2 + a_1^2(t) dx^2 + a_2^2(t) \left( dy^2 + s^2(y) dz^2 \right)
\; ,
\ee
where
\be
s(y)  \equiv \left\{
\begin{array}{cr}
 \sin y  & {\rm for\ Kantowski-Sachs,} \\
 y       & {\rm for\ LRS\ Bianchi\ I,} \\
 \sinh y & {\rm for\ Bianchi\ III.}
\end{array} \right.
\ee
The volume scale factor $V=a_1a_2^2$. If we introduce a conformally
invariant time coordinate $d\xi\equiv{dt\over a_2}={d\tt\over\ta_2}$ and
let $X\equiv\ta_1\ta_2$, we can write the Einstein equations as
\ba
X'' + 4kX
 & = & 8\pi G (\trho-\tp) X\ta_2^2 \\
\left({\ta_1'\over \ta_1}\right)' + {X'\over X}{\ta_1'\over \ta_1}
 & = & 4\pi G (\trho-\tp) \ta_2^2 \\
\left({X'\over X}\right)^2 - \left({\ta_1'\over \ta_1}\right)^2 + 4k
 & = & 8\pi G\trho+\hat{\rho} \ta_2^2 \; , \label{KSconstraint}
\ea
where $'\equiv {d/d\xi}$ and $k=+1,0,-1$ corresponds to
Kantowski-Sachs, Bianchi I or Bianchi III respectively, in analogy
with FRW models.

We have only been able to solve these equations analytically in the
presence of a stiff fluid plus the Brans-Dicke field. Even in the case
of radiation, where one can obtain both $\trho$ and $\hat{\rho}$ as
functions of $\ta_1$ and $\ta_2$, the resulting equations for $\ta_1$
and $\ta_2$ still cannot be integrated.
However, we see that a stiff fluid in the Einstein frame, like the
Brans-Dicke field, does not enter the first two equations and so (for
$\Gamma=0$) we can integrate both of these directly to give
\be
X = {\bar{A} \over 2} s(2[\xi-\xi_0]) \ ,
\ee
with the appropriate function $s(x)$ defined above and thus
\ba
\ta_1 & = & \ta_{1*} \tau^C \ , \nonumber \\
\ta_2 & = & \ta_{2*} {\tau^{1-C}\over 1+k\tau^2} \ , \\
\label{aBIIIKS}
\ea
where we have defined
\be
\tau \equiv \left\{
\begin{array}{cr} |\tan(\xi-\xi_0)|  & {\rm for\ Kantowski-Sachs,} \\
	      |\xi-\xi_0|  & {\rm for\ LRS\ Bianchi\ I,} \\
	      |\tanh(\xi-\xi_0)| & {\rm for\ Bianchi\ III.}
\end{array}
\right.
\ee
$\bar{A}$, $C$ and $\ta_{1*}$ are integration constants and
$\ta_{2*}=\bar{A}/\ta_{1*}$. The evolution of the two scale factors is
shown in Fig.(3).

The energy density of the stiff fluid in the Einstein frame is given
by Eq.~(\ref{trhochi}) where the constraint Eq.~(\ref{KSconstraint})
requires
\be
A^2 = {4\over3} \ta_{1*}^2 \ta_{2*}^2 (1-C^2) \; ,
\ee
and so clearly we must have $C^2<1$ in the presence of a stiff fluid
in the Einstein frame, or $C=\pm1$ in general relativistic
(i.e.~$\phi=$const) vacuum.

Note that we again obtain Kasner type solutions near the singularity
at $\xi\to\xi_0$, independent of the sign of $k$. The curvature
becomes irrelevant and we recover a
LRS subset of the Bianchi type~I solutions,
$a_i\propto|\eta-\eta_0|^{3c_i/2}$, where
$d\eta\equiv d\tt/(\ta_1\ta_2^2)^{1/3}$ and
\be
c_1 = {C \over 2-C} \qquad {\rm and} \qquad c_2=c_3={1-C \over 2-C} > 0 \; .
\ee
These $c_i$ obey the relations given in Eq.~(\ref{BIci}) with
\be
\Sigma^2 = \left( {2a_{1*}a_{2*}(1-2C)\over3} \right)^2 \; .
\ee
The area of the two-dimensional sub-space always vanishes as we
approach the singularity, while the remaining spatial dimension is
free to diverge (for $C<0$) or collapse ($C>0$). Thus the singularity
is linear or point-like.
In the Kantowski-Sachs case we find another singularity at
$\xi\to\xi_0+\pi/2$. This is like the initial singularity but
with $C\to-C$, and thus a point-like singularity is followed by a
linear singularity, or vice versa (unless $C=0$ in which case
$\ta_1$ remains constant throughout).

We can then integrate Eq.~(\ref{trhochi}) to obtain
\ba
\sqrt{16\pi G\over3} (\chi-\chi_0)
 & = & \int {Ad\xi\over X} \\
 & = & \sqrt{4(1-C^2)\over3} \ln \tau \; .
\ea
Thus $\chi$ must diverge both at early and late times in the
Kantowski-Sachs (or Bianchi type I) metric, but the expanding universe becomes
curvature dominated in the Bianchi type III metric with positive
spatial curvature and $\chi$ coasts to a fixed value
$\chi\to\sqrt{4(1-C^2)/3}$.

\subsubsection{LRS Bianchi type~IX}

To give an example of an anisotropic cosmology with closed spatial
hypersurfaces we consider a Bianchi type~IX metric
\cite{Nariai72,MatRyaTot73}, whose homogeneous
spatial hypersurfaces have volume $V=16\pi a_1a_2a_3$,
so the expansion
\be
\theta = {\dot{a_1}\over a_1} + {\dot{a_2}\over a_2} + {\dot{a_3}\over a_3}
\; .
\ee
When $\dot{a_1}/a_1=\dot{a_2}/a_2=\dot{a_3}/a_3$ we recover the closed
FRW model.

As the curvature terms in anisotropic models become more complicated,
our ability to give analytic solutions becomes more restricted. In the
case of Bianchi type~IX we can only give analytic solutions for the
locally rotationally symmetric (LRS) case (where $a_2=a_3$) plus stiff fluid
in the Einstein frame \cite{Bat+Coh72}, corresponding to vacuum or
stiff fluid solutions in scalar-tensor gravity. This will not show the
chaotic behavior of the more general Bianchi IX metric
\cite{Barrow82} nor the isotropizing effect of matter such as
isotropic radiation. Such issues would require a phase-space analysis and is
beyond the scope of this paper but is perhaps a topic worthy of a
investigation in its own right.

The equations of motion for the scale factors in the Einstein frame are then
\ba
{d\over d\tt}\left({\dot{\ta_1}\over\ta_1}\right)
 + \ttheta {\dot{\ta_1}\over\ta_1} = - {\ta_1^2\over 2\ta_2^4} \; ,\\
{d\over d\tt}\left({\dot{\ta_2}\over\ta_2}\right)
 + \ttheta {\dot{\ta_2}\over\ta_2} = {\ta_1^2-2\ta_2^2\over 2\ta_2^4} \; ,
\ea
plus the constraint equation
\be
\left({\dot{\ta_2}\over\ta_2}\right)
 + 2{\dot{\ta_1}\over\ta_1}{\dot{\ta_2}\over\ta_2}
 + {4\ta_1^2\ta_2^2-\ta_1^4 \over 4\ta_1^2\ta_2^4}
 = {3A^2 \over 4\ta_1^2\ta_2^4} \; .
\ee
where the only energy density is $\trho_\chi$ given in Eq.~(\ref{trhochi}).

Introducing the volume weighted time coordinate
$d\zeta=d\tt/(\ta_1\ta_2^2)$ and using the variables $x=4\ta_1^4$
and $y=4\ta_1^2\ta_2^2$, we can re-write the equations of motion as
\ba
2\left({x'\over x}\right)' + x & = & 0 \; , \\
2\left({y'\over y}\right)' + y & = & 0 \; ,
\ea
which can immediately be integrated to give
\ba
\ta_1^2 & = & {w_1 \over \cosh w_1(\zeta-\zeta_1)} \; , \nonumber \\
\ta_2^2 & = & {w_2^2 \cosh w_1(\zeta-\zeta_1)
	       \over w_1 \cosh^2 w_2(\zeta-\zeta_2)} \; ,
\label{aBIX}
\ea
as shown in Fig.(4),
where $\zeta_1$, $\zeta_2$, $w_1$ and $w_2$ are integration constants.
Clearly, $w_1$ must be positive and we can also take $w_2$ to be
positive without loss of generality.

The constraint equation requires
\be
3A^2 = 4w_2^2 - w_1^2 \; .
\ee
This is sufficient to ensure that $w_1\leq 2w_2$ and
both $\ta_2$ and $\ta_1$ approach zero as $\zeta\to\pm\infty$ (for
non-zero $w_1$ and $w_2$), the interval between this big bang and big crunch
taking only a finite proper time in the Einstein frame. Near these
singularities the scale factors evolve as power-laws, with respect
to proper time (or conformal time), and we recover another
one-parameter (for given $A$) subset of the Bianchi type~I solutions,
with
\be
c_1 = {w_1\over 4w_2-w_1} \; , \qquad
c_2 = c_3 = {2w_1-w_1\over 4w_2-w_1} \; .
\ee
The initial (and final) shear then approaches $\tsigma^2\to
3\Sigma^2/4\ta^6$ with $\Sigma^2\equiv 4(w_2-w_1)^2/9$.

Note that the equation for the scalar field energy density,
Eq.~(\ref{trhochi}), shows that $\chi$ is just proportional to the
volume weighted time, $\zeta$, and so we have
\be
\sqrt{16\pi G\over 3} (\chi-\chi_0) = \pm A \zeta \; ,
\ee
which must also diverge as we approach both the initial and final
singularity.

\section{Conclusions}

We have derived a number of new exact solutions for anisotropic
cosmologies in scalar-tensor gravity theories where previous results
were in general restricted to isotropic models
\cite{Barrow93,Dam+Nor93,Bar+Mim94,Mim+Wan95}
or the particular case of Brans-Dicke gravity
\cite{Nariai72,MatRyaTot73,Bel+Kha73,Rub+Fin75,Ruban77,LP}.
In the original, or Jordan, frame the non-minimal coupling of the
Brans-Dicke field introduces an effective anisotropic pressure
proportional to the existing shear.  This can significantly modify the
evolution even in vacuum, especially near a singularity. It is
possible for the Brans-Dicke field to reverse the collapse as we
approach an anisotropic singularity and lead to a non-singular,
isotropic expanding universe. In the Brans-Dicke gravity theory the deviation
from general relativistic behavior is strictly limited due to present
day observational limits on the Brans-Dicke parameter $\omega$ \cite{Will93}.
However in more general scalar-tensor gravity the present day value
need not constrain the value of $\omega$ near an initial singularity.

We have emphasized the importance of using the conformally related
Einstein metric where the Brans-Dicke field is minimally coupled with
respect to the metric and so the usual general relativistic results
hold good. Here, for instance, there is no anisotropic pressure due to
the Brans-Dicke field. Non-Einstein behavior only appears in the
transformation back to the Jordan frame.

The analytic complexity of the evolution in the Einstein frame lies in
the interaction introduced directly between the Brans-Dicke field and
ordinary matter. Except in the particular cases of vacuum, radiation
or a stiff fluid, this forces us to restrict our results for general
barotropic fluids [$p=(\gamma-1)\rho$] to the spatially flat Bianchi
type~I metric. Here the solutions for general $\omega(\phi)$ theories
are characterized by a generating function $g(\xi)$, where $\xi$ is a
time coordinate. Brans-Dicke gravity corresponds to the case where
$g\propto(\xi+\xi_0)^2$.

The shear vanishes relative to the density of matter and/or the
Brans-Dicke field (i.e., the models isotropize) whenever
$|\xi|\to\infty$. This requires simply that $g$ remains positive,
or equivalently that the Brans-Dicke parameter
$\omega\geq2(\gamma-5/3)/(2-\gamma)^2$, as $|\xi|\to\infty$. If
$g<\alpha\xi^2$ for any constant $\alpha>0$ as $\xi\to\infty$ then the
dynamics become dominated by the Brans-Dicke field density at late times.
Conversely, if $g>\alpha\xi^2$ then the models become dominated by the
barotropic matter and we recover the general relativistic behavior
where the Brans-Dicke field $\phi\to$const and $\omega\to\infty$.
Brans-Dicke gravity is seen to correspond to the particular case where
the relative densities of the Brans-Dicke field and ordinary matter
remain proportional.

Singularities in the Jordan frame (where the volume scale factor
vanishes) occur only when the effect of ordinary matter
(with $p<\rho$) becomes negligible compared with the shear and
Brans-Dicke field energy density. Thus it is sufficient to consider
only the vacuum or stiff fluid ($p=\rho$) models to discuss the
approach to the singularity.

We have shown that in the presence only of radiation and a stiff fluid
(equivalent to the short and long wavelength modes of massless fields
\cite{Mim+Wan95}), or in vacuum, the evolution of the scale factor in
the Einstein frame is independent of the form of $\omega(\phi)$ and
corresponds exactly to the standard general relativistic evolution
with radiation and a stiff fluid. This enables us to give results for
arbitrary $\omega(\phi)$ theories in Bianchi type~I, III, V and LRS
type~IX and Kantowski-Sachs metrics. At singularities in the Einstein
frame these all approach a Bianchi type~I solution as the spatial
curvature becomes negligible. At this point the energy density of the
scalar field in the Einstein frame, $\chi$, must diverge. The relation
between this field and the original Brans-Dicke field depends on the
form of $\omega(\phi)$, but in any case a necessary condition for the
divergence of $\chi$ is that $\ln(\phi)$ diverges or
$\omega\to\infty$.  However the simplest general relativistic limit,
considered above as a late-time attractor in Bianchi type~I, in vacuum
or in the presence solely of radiation or stiff fluid, is not
sufficient to lead to a divergence of $\chi$ and does not act as an
attractor solution at the initial singularity. On the other hand
$\phi\to0$ is an attractor which can lead to non-singular behavior
in the Jordan frame.

Thus while many scalar-tensor gravity theories do approach general
relativity at late times, where the role of other barotropic matter is
dominant and anisotropy vanishes, they may nonetheless give markedly
non-Einstein behavior near the initial big bang where anisotropy is
important.

\section*{Acknowledgments}

DW is a PPARC Research Fellow. JPM would like to thank Mariano Quir\'os
at the Instituto de Estructura de la Materia, CSIC in Madrid and
Professor Roger Tayler and the Astronomy Centre at the University of
Sussex for their hospitality while part of this work was carried out.
The authors acknowledge use of the Starlink computer system at Sussex
and would like to thank John Barrow, Alberto Casas, Jesus Moreno and
Jon Yearsley for helpful discussions.

%%%%%%%%%%%%%%%%%%%%%%%%%%%%%%%%%%%%%%%%%%%%%%%%%%%%%%%%%%%%%%%%%%%%%%%%%

%%%%%%%%%%%%%%%%%%%%%%%%%%%%%%%%%%%%%%%%%%%%%%%%%%%%%%%%%%%%%%%%%%%%%

\section*{Figure Captions}

\noindent
{\bf Figure 1\\}
The Brans-Dicke field $\phi$, as a function of the minimally coupled
field in the Einstein frame, $\chi$. In Brans-Dicke gravity (with
$\omega=1$) $\phi(\chi)$ is given by solid lines (for two choices of
sign of $\beta_0$) in vacuum or radiation ($A^2=2$, $M=0$) or by dotted
line in presence of a stiff fluid ($M=0.2$).  For $\omega(\phi)$ given
in Eq.~(\ref{defBDGR}), with $\phi_0=0.8$, the dashed line gives the
$\phi(\chi)$ for $M=0$, or dotted-dashed line for $M=0.2$.

\vspace{0.1in}
\noindent
{\bf Figure 2\\}
Evolution of the three independent scale factors $\ta_i$, in the
Einstein frame, in Bianchi type~I metric as given in Eq.~(\ref{aBI}) with
$A^2=2$ and with radiation (solid line, $\Gamma=2$) or without (dashed
line, $\Gamma=0$). We have chosen $c_1=-1/4$, $c_2=1/2$, and $c_3=3/4$
in both cases.

\vspace{0.1in}
\noindent
{\bf Figure 3\\}
Evolution with respect to time coordinate $\xi$ of the two independent
scale factors $\ta_1$ and $\ta_2$, in the Einstein frame, in
Kantowski-Sachs (solid line) or Bianchi type~III metric (dashed) as
given in Eq.~(\ref{aBIIIKS}). We have taken $C=0.3$.

\vspace{0.1in}
\noindent
{\bf Figure 4\\}
Evolution with respect to time coordinate $\zeta$ of two independent
scale factors $\ta_1$ and $\ta_2$, in Einstein frame, in LRS Bianchi
type~IX metric as given in Eq.~(\ref{aBIX}).
We have taken $A^2=2$ and $w_1=5$.

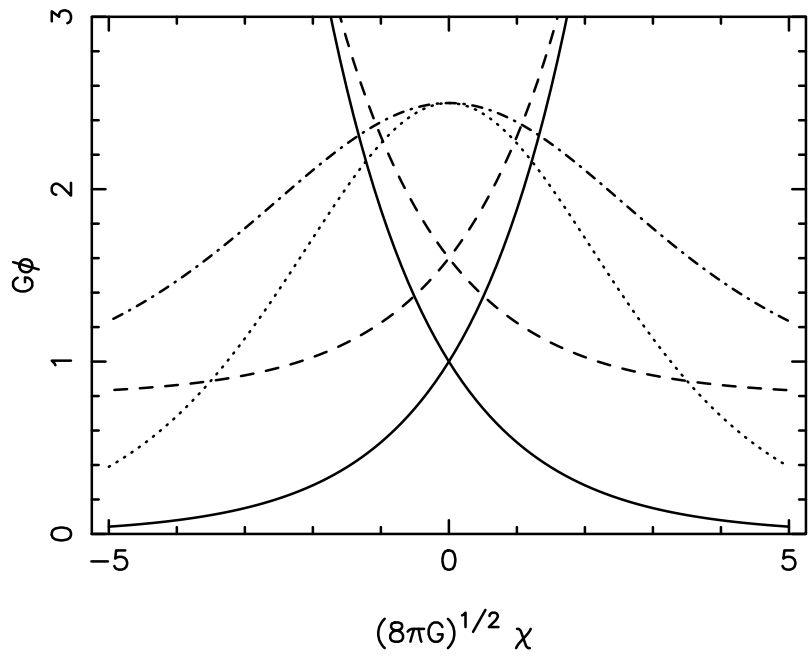
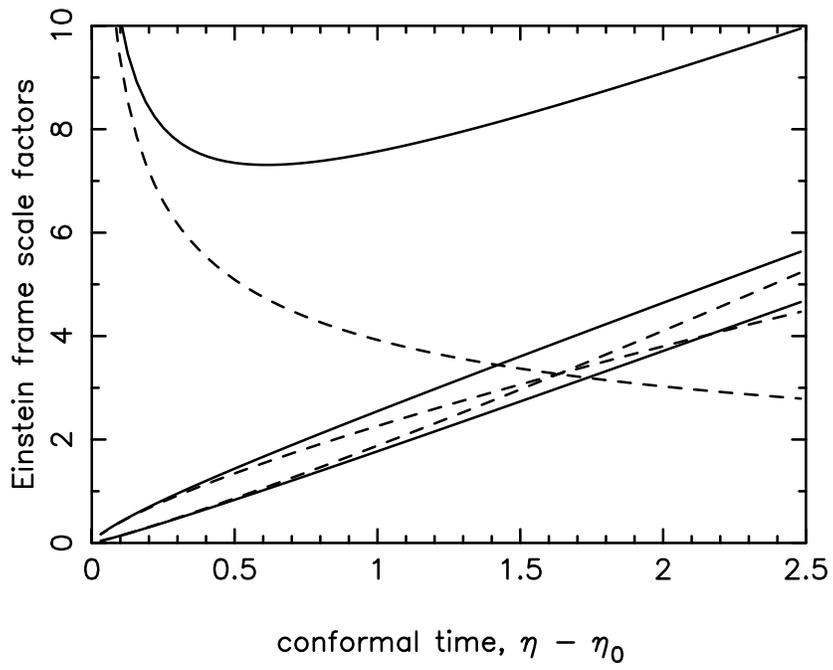
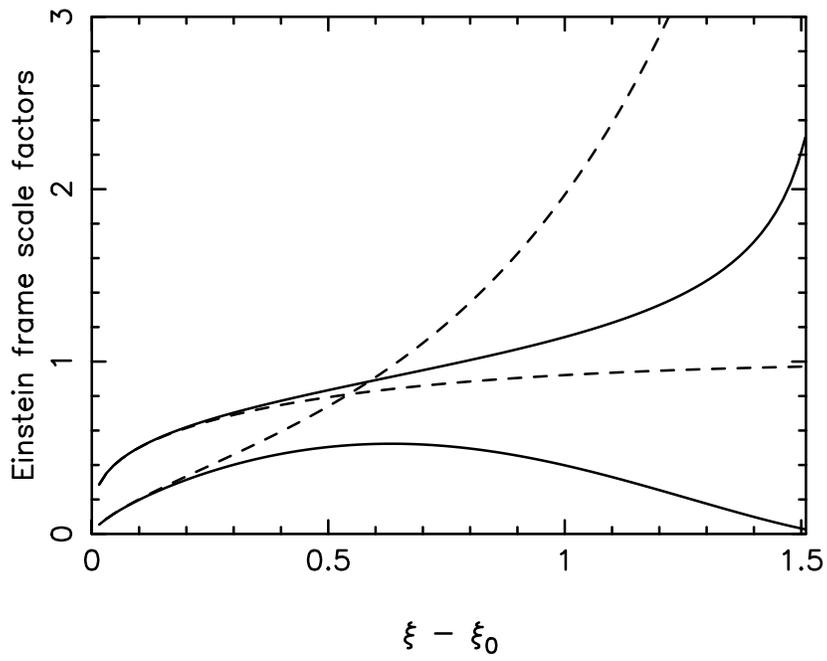
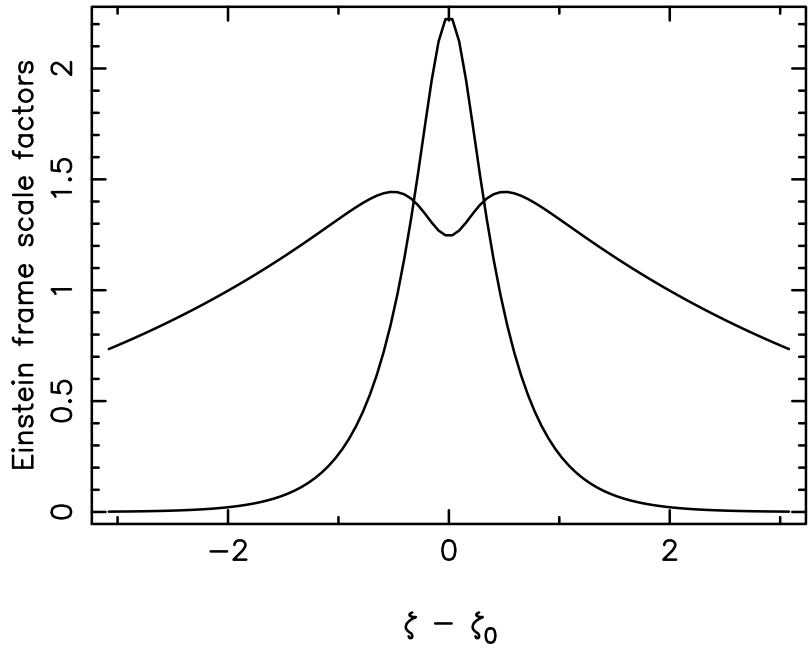
\end{document}